\documentclass[sigconf]{acmart}

\usepackage{soul}
\renewcommand\footnotetextcopyrightpermission[1]{} 
\usepackage[utf8]{inputenc}
\usepackage{comment}
\usepackage{multirow}
\usepackage{pifont}
\usepackage{xifthen}
\usepackage[underline=false]{pgf-umlsd}
\usepackage{tikz}
\usepackage{algorithm}
\usepackage[noend]{algpseudocode}
\usepackage{algpseudocode}
\newcommand{\formatcomment}[1]{\scriptsize\textcolor{blue!25!black}{\texttt{#1}}}

\algrenewcommand{\algorithmiccomment}[1]{\hfill\parbox{5.2cm}{\formatcomment{//\,#1}}}
\algrenewcommand\algorithmicprocedure{\textbf{algorithm}}
\algnewcommand\algorithmicforeach{\textbf{for each}}
\algdef{S}[FOR]{ForEach}[1]{\algorithmicforeach\ #1\ \algorithmicdo}
\usepackage{enumitem}
\usepackage[n ,advantage, operators, sets , adversary, landau, probability, notions, logic, ff , mm, primitives, events, complexity, asymptotics, keys]{cryptocode}
\usepackage{booktabs} 
\usepackage{comment}
\usepackage[normalem]{ulem} 
 \newcommand{\superscript}[1]{\ensuremath{{}^{\textrm{\scriptsize #1}}}}
 
 \newcommand{\mntext}[1]{\colorbox{SkyBlue}{\begin{color}{black}#1\end{color}}}
 \newcommand{\mn}[2][]{{\tiny\superscript{\mntext{\arabic{mn}}}}\marginpar{\scriptsize{
 			\ifthenelse{\isempty{#1}}
 			{\mntext{\parbox{0.95\marginparwidth}{\superscript{\arabic{mn}}~\raggedright{#2}}}}
 			{\mntext{\parbox{0.95\marginparwidth}{\superscript{\arabic{mn}}#1 says~:~\raggedright{#2}}}}
 		}}\stepcounter{mn}}

\usepackage{mathtools}
\usepackage{graphicx}  
\usepackage{xcolor,colortbl}
\usepackage{tikz}
\usepackage{pgfplots}
\usepackage{subfigure}
\usetikzlibrary{decorations.pathreplacing,calc}

\theoremstyle{definition}
\newtheorem{definition}{Definition}[section]
\newtheorem{theorem}{Theorem}[section]
\newtheorem{corollary}{Corollary}[theorem]

\theoremstyle{remark}

\title{
On Privacy and Confidentiality of Communications in Organizational Graphs}
\date{}
\usepackage{natbib}

\author{Masoumeh Shafieinejad}
\affiliation{%
  \institution{University of Waterloo}
  \city{Waterloo}
  \country{Canada}}
\email{masoumeh@uwaterloo.ca}

\author{Huseyin Inan}
\affiliation{%
  \institution{Microsoft Research}
  \city{Redmond}
  \country{United States}}
\email{Huseyin.Inan@microsoft.com p}

\author{Marcello Hasegawa}
\affiliation{%
  \institution{Microsoft Corporation}
  \city{Redmond}
  \country{United States}}
\email{marcellh@microsoft.com}

\author{Robert Sim}
\affiliation{%
  \institution{Microsoft Research}
  \city{Redmond}
  \country{United States}}
\email{rsim@microsoft.com}

\begin{document}
\pagestyle{plain} 
\begin{abstract} 
Machine learned models trained on organizational communication data, such as  emails in an enterprise, carry unique risks of breaching confidentiality, even if the model is intended only for internal use.  This work shows how confidentiality is distinct from privacy in an enterprise context, and aims to formulate an approach to preserving confidentiality while leveraging principles from differential privacy. The goal is to perform machine learning tasks, such as learning a language model or performing topic analysis, using interpersonal communications in the organization, while not learning about confidential information shared in the organization. Works that apply differential privacy techniques to natural language processing tasks usually assume independently distributed data, and overlook potential correlation among the records. Ignoring this correlation results in a fictional promise of privacy. Naively extending differential privacy techniques to focus on group privacy instead of record-level privacy is a straightforward approach to mitigate this issue. This approach, although providing a more realistic privacy-guarantee, is over-cautious and severely impacts model utility. We show this gap between these two extreme measures of privacy over two language tasks, and introduce a middle-ground solution. We propose a model that captures the correlation in the social network graph, and incorporates this correlation in the privacy calculations through Pufferfish privacy principles.
\end{abstract}

\maketitle
\section{Introduction}
A number of applications in natural language understanding rely on language models~\cite{chen2019gmail,adam2020ai}.
To enable such models it is necessary to process training data that best represents the target application.
Such tasks become especially sensitive in the setting of organizational communication, where organizations, individuals, or communities may share secret data, and preserving confidentiality is of utmost importance. 
Organizational communication often presents a complex underlying structure of interactions which is well modeled by a social graph. Data privacy in graphs and social graphs have been addressed by a number of previous works, for example~\cite{189,134,43,kearns2016private,gao2017preserving,zhu2017differentially,Karwa2014}, 
where most of the works based on differentially private approaches model individuals as nodes and exchanged messages as edges. These proposed methods provide either node level privacy guarantees or edge level privacy guarantees. In the context of organizational communication we define node level guarantees as individual privacy and edge level guarantees as confidentiality. That is, confidentiality involves protecting information that is shared between two or more individuals in the organization. In this work we set aside questions of individual privacy and examine problems in ensuring confidentiality in organizational communication.  The goal is to enable the production of ML models for an enterprise, without compromising potentially sensitive business secrets. 

A limitation of individual privacy is that in an organizational context, this approach can produce models with lower than optimal utility.  To understand this, suppose the CEO of a company communicates regularly with messages addressed to all employees. From the organization's perspective this information is public to everyone, and yet a differentially private mechanism acting at a node level will be forced to mask out the presence of these messages as they originate from a single node.
This would be necessary to satisfy the requirement that a differentially private model should be invariant to the presence or absence of any single node. 

Instead we consider a model that affords edge-level privacy.  In this case the CEO can broadcast to all employees, and enable this correspondence to be included in training of the model, since the information is replicated across many edges, but she can still protect private 1:1 correspondence with her CFO, for example, thus preserving confidentiality.

However, the situation is not so straightforward.  
In the case of social graphs, often the properties of an edge can be inferred by the properties of other nearby edges. Such a case does not fall under the framework of differential privacy, where the theoretical guarantees are based on presence/absence of individual elements, without taking into account any effect on other elements. 
One approach to address this problem is via group differential privacy. Group privacy assumes that all edges in a group of participants are fully correlated, and queries against the graph must be invariant to the presence or absence of the entire group.  This framing can significantly impact the accuracy of the model or query by being over-protective of edge-edge relationships. Furthermore, it requires an explicit description of what constitutes a group in the organization.  

We address the issues presented by edge correlation by employing a generalized version of differential privacy called Pufferfish \cite{song2017}. In Pufferfish a set $\mathcal{S}_{\textnormal{pairs}} \subseteq \mathcal{S} \times \mathcal{S}$ of complementary secret pairs is defined, and privacy is provided by ensuring the secret pairs are indistinguishable, for any data distribution (capturing the correlation among the records) known to the adversary. Through this requirement the correlation problem can be addressed while allowing utility to be preserved. In addition, the non-independence between edges also affords us a simple model for confidentiality, namely that information passing between neighboring edges is more likely to be confidential than information that is randomly distributed in the social graph.  From this perspective, we propose a privacy model that accounts for information dependence between edges in the graph and define a notion of what constitutes an edge's neighborhood. In this work we are limited to L-Lipschitz queries which are sufficiently broad to cover for counting and frequency queries.

For the purposes of this work we adapt the \emph{Attribute Disclosure Attack} defined in~\cite{Beigi2020}:
\begin{definition} [Attribute Disclosure Attack] Given T=(G,A,B), which is a snapshot of an organization with a social graph $G=(V,E)$, where $V$ is the set of individuals and $E$ demonstrates the correspondence between them, correspondent behavior $A$ and attribute information $B$, the attribute disclosure attack is used to infer the attributes $a_e$ for all $e\in E_t$ where $E_t$ is a list of targeted edges. For each $e \in E_t$, we have information about the correspondence between the individuals linked by $e$.
\end{definition}

This definition deviates from that proposed in~\cite{Beigi2020} in that it focuses on disclosing information about correspondence between users. That is, the goal of the attack is to leak information about what is communicated along the edges of the graph.

\subsection{Contributions}
The contributions of this work are as follows, it:
\begin{itemize}
\item Presents a complete formulation of \emph{confidentiality} protection in closed social networks such as an enterprise or organization, including accounting for neighborhood correlation, 
\item Models predicting neighborhood correlations assuming varying degrees of attacker knowledge,
\item Shows the gap between two extreme measures of privacy, record-level privacy and group privacy, over two language tasks, and
\item Provides empirical results of applying our work to statistical query tasks~\cite{Gopi2020,xu2013differentially} in a real enterprise graph, leaving extensions to other language modeling tasks for future work. 

\end{itemize}

\subsection{Related Work}
Existing privacy approaches for social graphs use different techniques and mechanisms \cite{Beigi2020}. These techniques are categorized into three main categories: i) grouping-based approaches, ii) edge manipulation algorithms, and iii) differential-privacy-based techniques. Grouping-based approaches include $k$-anonymity-based approaches \cite{43,115,189,196,199} and cluster-based techniques \cite{31,70,114,134,174}. $K$-anonymity is among the first techniques proposed for protecting the privacy of datasets and aims to anonymize each user/node in the graph so that it is indistinguishable from at least $k-1$ other users. Machanacajjhala et al.~\cite{Machanavajjhala2007} showed that a $k$-anonymous solutions still have severe privacy problems when the sensitive attributes lack diversity, or when the adversary has access to background knowledge. Clustering-based approaches group users and edges and are limited to the applications where only the density and size of the cluster is revealed, so that individual attributes are protected. Edge manipulation algorithms utilize edge-based strategies such as random edge adding/deleting and random edge switching \cite{188,18}. The edge perturbation algorithm can use random-walk-based techniques \cite{116,134} as well. As the most recent category, differential privacy provides a strong privacy guarantee that the risk of user’s privacy leakage does not increase as a result of participating in a database \cite{52}. A common way of achieving differential privacy is by introducing random noise through the Laplace mechanism (for numerical attributes) or Exponential mechanism (for non-numerical attributes) to the query answers \cite{52}. 

The differentially private proposals for social graphs fall into two categories: edge-level differential privacy, and node-level differential privacy. In edge-level differential privacy \cite{Nissim2007,Sala2017,Karwa2014,Hay2009}, the result of the analysis does not change by adding or removing an edge. Node-level privacy however \cite{Kasiviswanathan2013,Blocki2013,Chen2013,Raskhodnikova2016,Chen2014,Ghosh2017}, is resilient against adding or removing a node, therefore it is more difficult to achieve. Note that differential privacy assumes independence among instances in the dataset. It has been shown that the dependency between instances affects the robustness of differential privacy guarantees to de-anonymization  \cite{liu2016dependence,92}. The naive approach to address this issue to use group-level differential privacy, where the group of dependant records/nodes are considered as one instance \cite{Chen2014,Ghosh2017}. The transition to group privacy sacrifices utility even more, and motivates our work. We investigate privacy for correlated data, as an alternative to group-privacy, to prevent leakage when the records are not independent \cite{journal-1,song2017}. To do so, we use Pufferfish privacy \cite{journal-1}, and propose a mechanism to achieve this privacy inspired by Song et al.'s work \cite{song2017}.

\nocite{zhu2017differentially,gao2017preserving,kearns2016private,yang2015}

\section{Preliminaries}
In this section, we introduce the notations and provide the definitions for the concepts used throughout the paper.

\subsection{Differential Privacy}
Differential privacy (DP) is a mathematical framework that offers strong and robust guarantees at protecting user privacy under the release of a query function calculated over a statistical database \cite{proceedings-1}. The main idea to achieve DP is perturbing the query function by the introduction of random noise generated according to a carefully chosen distribution. We formally define the notion of $\epsilon$-differential privacy in the following:
\begin{definition}[$\epsilon$-Differential Privacy \cite{proceedings-1}]
A randomized algorithm $\mathcal{M}$ is said to provide $\epsilon$-differential privacy if for any two databases $D, D'$ differing in only a single entry, and for any set $S \subseteq \textnormal{Range}(M)$,
\begin{align*}
    \dfrac{\Pr(\mathcal{M}(D) \in S)}{\Pr(\mathcal{M}(D') \in S)} \leq \exp(\epsilon)
\end{align*}
the probability taken over the randomness of the algorithm $\mathcal{M}$.
\end{definition}

A typical way to achieve $\epsilon$-differential privacy is to apply the Laplace mechanism as shown in \cite{proceedings-2}. The main idea is to apply noise to the output of a query for the sake of perturbation and the amount of noise depends on the global sensitivity of the query and the privacy budget $\epsilon$. Let us first define the global sensitivity. 
\begin{definition}[Global sensitivity]
Let $d$ be a positive integer and $\mathcal{D}$ be a collection of datasets. 
For any function $f : \mathcal{D} \rightarrow \mathbb{R}^d$, the global sensitivity of f, denoted by $\Delta f$, is defined by 
\begin{align*}
   \Delta f = \max \limits_{D_1, D_2} || f(D_1) - f(D_2) ||_1
\end{align*}
where $D_1$ and $D_2$ are datasets in $\mathcal{D}$ differing in at most one element and $||\cdot||_1$ denotes the $\ell_{1}$ norm.
\end{definition}

\begin{theorem}
Let $d$ be a positive integer and $\mathcal{D}$ be a collection of datasets. 
For any $f : \mathcal{D} \rightarrow \mathbb{R}^d$, the randomized mechanism $\mathcal{M}$
\begin{align*}
    \mathcal{M}(\mathcal{D}) = f(\mathcal{D}) + \textnormal{Laplace}(0, \Delta f/\epsilon)
\end{align*}
satisfies $\epsilon$-differential privacy. 
\end{theorem}

Global sensitivity is a bound on the maximum effect of any element in the dataset, which will be helpful to provide privacy to \textit{all} elements in the dataset. Based on this, for a given dataset $D$, a function $f$, and the privacy parameter $\epsilon$, the Laplace mechanism adds $\textnormal{Laplace}(0, \lambda)$ noise to the output of $f$ where the parameter $\lambda$ is determined by both $\Delta f$ and $\epsilon$.

Note that in the definition of $\epsilon$-differential privacy, the guarantee holds for a single data entry. However, by applying the composability property of differential privacy \cite{mcsherry09}, the setting can  be extended to multiple data entries. If the goal is to protect a group, this can be achieved by setting $\epsilon$ to $\epsilon/k$ for any $k \in \mathbb{N}$ where $k$ represents the size of the group. In this case, all groups of size $k$ are $\epsilon$-differentially private protected. We summarize this by formally defining group differential privacy in the following corollary:
\begin{corollary}[Group Differential Privacy]
A randomized algorithm $\mathcal{M}$ is said to provide $\epsilon$-differential privacy for all groups of size $k$ if for any two databases $D, D'$ differing in at most $k$ entries, and for any set $S \subseteq \textnormal{Range}(M)$,
\begin{align*}
    \dfrac{\Pr(\mathcal{M}(D) \in S)}{\Pr(\mathcal{M}(D') \in S)} \leq \exp(\epsilon)
\end{align*}
where the probability is taken over the randomness of the algorithm $\mathcal{M}$. Any $\epsilon/k$-differentially private algorithm is $\epsilon$-differentially private for all groups of size $k$.
\label{group_DP}
\end{corollary}

We point out that the scaling of the noise is inversely proportional to the privacy budget $\epsilon$. Therefore, setting $\epsilon$ to $\epsilon/k$ will in turn change the noise level from $\Delta f/\epsilon$ to $k \cdot \Delta f/\epsilon$, which may significantly decrease the utility of the query. However, this is the price to pay to obtain stronger privacy guarantees with group differential privacy.  
 
\subsection{Pufferfish Privacy}
Differential privacy provides robust guarantees for a wide range of database queries.  For our scenario, it is useful to consider Pufferfish privacy~\cite{journal-1}, which is a Bayesian privacy framework providing rigorous privacy guarantees against many types of attackers. An advantage of the Pufferfish framework is that a domain expert can develop rigorous privacy definitions for their data sharing needs without holding an expertise in privacy. 
This is achieved by specifying three components in the Pufferfish privacy framework: a set $\mathcal{S}$ of potential secrets, a set $\mathcal{S}_{\textnormal{pairs}} \subseteq \mathcal{S} \times \mathcal{S}$ of discriminative secret pairs, and a collection of data distributions $\Theta$. The Pufferfish framework provides a rich class of privacy definitions based on the components specified by a domain expert. We formally define the framework in the following based on \cite{journal-1}.
\begin{definition}[Pufferfish Privacy]
A randomized algorithm $\mathcal{M}$ is said to provide $\epsilon$-Pufferfish privacy for a domain $(\mathcal{S}, \mathcal{S}_{\textnormal{pairs}}, \Theta)$
if for all distributions $\theta \in \Theta$, for all secret pairs $(s_i, s_j) \in \mathcal{S}_{\textnormal{pairs}}$, and for all possible outputs $w \in \textnormal{Range}(\mathcal{M})$ it satisfies
\begin{align*}
    \left\lvert \dfrac{\Pr_{\mathcal{M}, \theta}(\mathcal{M}(\mathcal{D}) = w | s_i, \theta)}{\Pr_{\mathcal{M}, \theta}(\mathcal{M}(\mathcal{D}) = w | s_j, \theta)} \right\rvert \leq \exp(\epsilon)
\end{align*}
where $\mathcal{D}$ is drawn from the distribution $\theta$ and $s_i$ and $s_j$ are such that $\Pr(s_i | \theta) \neq 0$ and $\Pr(s_j | \theta) \neq 0$.
\end{definition}

We note that there is an additional source of randomness in the definition of Pufferfish privacy. The dataset $\mathcal{D}$ is itself a random variable that is drawn from a distribution $\theta \in \Theta$.

In words, a domain expert constructs the set $S$ for the potential secrets that are desired to be hidden (e.g. private data of an individual). $\mathcal{S}_{\textnormal{pairs}}$ is simply the pair of such potential secrets that we would like to guarantee are indistinguishable in evaluating $\mathcal{M}$. Finally, $\Theta$ is a collection of distributions where each probability distribution $\theta \in \Theta$ corresponds to an attacker to be protected against.  $\Theta$ can be selected based on the fine grain of how data can be plausibly generated and it also reflects the attackers' beliefs in how the data were generated (incorporating any background knowledge and side information). The whole process gives the domain expert flexibility to  customize privacy to the specific set of secrets and data generation scenarios that are typical in their domain.

We further point out that Pufferfish privacy provides a general framework in the sense that it covers $\epsilon$-differential privacy as an instantiation for a particular choice of domain $(\mathcal{S}, \mathcal{S}_{\textnormal{pairs}}, \Theta)$ (see Theorem 6.1 in \cite{journal-1}).

\subsection{Wasserstein Mechanism}
While there is no efficient general mechanism that applies to any Pufferfish instantiation, there are a number of mechanisms for specific Pufferfish instantiations \cite{journal-1, Blowfish}. For general Pufferfish instantiation, Song et al.~\cite{song2017} introduce a mechanism that achieves Pufferfish privacy, but does not satisfy efficiency in its original form. We introduce their base mechanism here. Later in Section \ref{PrivMech}, we present our adjustments to their mechanism that makes it efficient to utilize for our use case of enterprise communications. 

The main idea of the mechanism in \cite{song2017} is similar to the Laplace mechanism in differential privacy. Instead of adding noise based on the global sensitivity $\Delta f$ in differential privacy, Song et al. use the distributions $\Pr(f(\mathcal{D})| s_i, \theta)$ and $\Pr(f(\mathcal{D})| s_j, \theta)$ in the Pufferfish framework, propose a metric quantifying the worst case distance between these two distributions, and inject noise proportional to this distance. They find that the $\infty$-Wasserstein distance is the right choice for this purpose.
\begin{definition}[$\infty$-Wasserstein distance]
Let $\mu$, $\nu$ be two probability distributions on $\mathbb{R}$ and $\tau(\mu,\nu)$ denote the set of all joint distributions with marginals $\mu$ and $\nu$. The $\infty$-Wasserstein distance between $\mu$ and $\nu$ is defined as
\begin{align*}
    W_{\infty}(\mu, \nu) = \inf \limits_{\gamma \in \tau (\mu, \nu)} \max \limits_{(x,y) \in \textnormal{support}(\gamma)} \mid x-y \mid.
\end{align*}
\end{definition}
Intuitively, $W_{\infty}$ measures the maximal distance that any probability mass moves while transforming $\mu$ to $\nu$ in the most optimal way possible. 
$W_{\infty}$ is related to the well-known Earth Mover's Distance in that it accounts for the maximal shift in probability over the domain of $\tau$ but not the amount of mass in the shift \cite{hitchcock41}. 

Based on the $\infty$-Wasserstein distance, the Wasserstein mechanism calculates the maximum over $(s_i, s_j) \in \mathcal{S}_{\textnormal{pairs}}$ and $\theta \in \Theta$, analogous to $\Delta f$, and applies the Laplace noise proportional to the maximum $\infty$-Wasserstein distance. It is proven in \cite{song2017} (see Theorem 3.2) that this mechanism yields $\epsilon$-Pufferfish privacy.
\begin{theorem}[Wasserstein mechanism]
Let $(\mathcal{S}, \mathcal{S}_{\textnormal{pairs}}, \Theta)$ be a domain. For any function $f : \mathcal{D} \rightarrow \mathbb{R}$ the randomized mechanism $\mathcal{M}$
\begin{align*}
    \mathcal{M}(\mathcal{D}) = f(\mathcal{D}) + \textnormal{Laplace}(0, W/\epsilon)
\end{align*}
where $W = \sup_{(s_i, s_j) \in \mathcal{S}_{\textnormal{pairs}}, \theta \in \Theta} W_{\infty}(\mu_{i\theta}, \nu_{j\theta})$ for $\mu_{i, \theta} = \Pr(f(\mathcal{D})| s_i, \theta)$ and $\nu_{j, \theta} = \Pr(f(\mathcal{D})| s_j, \theta)$ satisfies $\epsilon$-Pufferfish privacy. 
\end{theorem}



\subsection{Markov Quilt Mechanism} \label{MarkovQuilt}
The Wasserstein mechanism can be quite expensive in terms of computational complexity, as it requires modeling the effects of varying all complementary secret pairs on the function $f$.  \cite{song2017} introduces the Markov Quilt mechanism for the special case where the dependence inside a dataset can be described by a Bayesian network, which fits to our setting of interest. In the case where the dependence is most effective in the ``local'' neighborhood, the amount of noise can be calibrated with respect to the size of this neighborhood. To this end, max-influence of a variable $\mathcal{D}_i$ on a set of variables $\mathcal{D}_A$ under a distribution class $\Theta$ is defined as 
{\small $$e_{\Theta}(\mathcal{D}_A | \mathcal{D}_i) = \sup \limits_{\theta \in \Theta} \max \limits_{a, b \in \mathcal{X}} \max \limits_{d_A \in \mathcal{X}^{|\mathcal{D}_A|}} \log \frac{\Pr(\mathcal{D}_A = d_A | \mathcal{D}_i = a, \theta)}{\Pr(\mathcal{D}_A = d_A | \mathcal{D}_i = b, \theta)}$$}
where $\mathcal{X}$ denotes the range of each $\mathcal{D}_i$.

In terms of privacy it is an advantage that the dependence stays as ``local'' as possible if one can find a large set $\mathcal{D}_A$ such that $\mathcal{D}_i$ has low max-influence on $\mathcal{D}_A$. Especially if one can claim certain conditional independence from a variable towards some part of the dataset it can also simplify the calibration of the noise. The following notion is helpful to show what is described here. 
\begin{definition}[Markov Quilt]
A set of variables $\mathcal{D}_Q$ in a dataset is a Markov Quilt for a variable $\mathcal{D}_i$ if there exists a set $\mathcal{D}_i \in \mathcal{D}_N$ such that $\mathcal{D} = \mathcal{D}_N \cup \mathcal{D}_Q \cup \mathcal{D}_R$ and $\mathcal{D}_i$ is conditionally independent from $\mathcal{D}_R$ given $\mathcal{D}_Q$, i.e. $\Pr(\mathcal{D}_R | \mathcal{D}_Q, \mathcal{D}_i) = \Pr(\mathcal{D}_R | \mathcal{D}_Q)$. 
\end{definition}

In this formulation \cite{song2017} choose the subscripts $N$ and $R$ to represent ``nearby'' and ``remote'' nodes in the Bayesian network, respectively, with the $Q$ (quilt) nodes separating them and establishing conditional independence.

Based on this notion \cite{song2017} introduces the Markov Quilt mechanism that protects a variable $\mathcal{D}_i$ by adding Laplace noise to a L-Lipschitz query $f$ with scale parameter $L \times |\mathcal{D}_N|/(\epsilon - \delta)$ where $\delta$ is an upper bound on the max-influence 
of $\mathcal{D}_i$ on $\mathcal{D}_Q$. We note that the effect of $\mathcal{D}_i$ is obscured with noise whose amount is based on the cardinality of the local variables ($|\mathcal{D}_N|$) and a correction term to account for the effect of the distant variables ($\delta$). Naturally, the privacy of all variables can be protected by adding noise with the maximum scale parameter over all variables $\mathcal{D}_i \in \mathcal{D}$. It is shown in \cite{song2017} (see Theorem 4.3) that this mechanism yields $\epsilon$-Pufferfish privacy. 
It is also proven that Markov Quilt mechanism satisfies sequential composition (see Theorem 4.4 in \cite{song2017}).

\begin{theorem}[Markov Quilt Mechanism]
Let $(\mathcal{S}, \mathcal{S}_{\textnormal{pairs}}, \Theta)$ be a domain. For any L-Lipschitz function $f$ if each $\mathcal{D}_i \in \mathcal{D}$ has the trivial quilt $\mathcal{D}_Q = \emptyset$ (with $\mathcal{D}_N=\mathcal{D}$, $\mathcal{D}_R = \emptyset$), then the Markov Quilt Mechanism provides $\epsilon$-Pufferfish privacy.
\end{theorem}

\section{Problem Definition}
\begin{figure}[ht]
    \centering
    \includegraphics[scale=0.35]{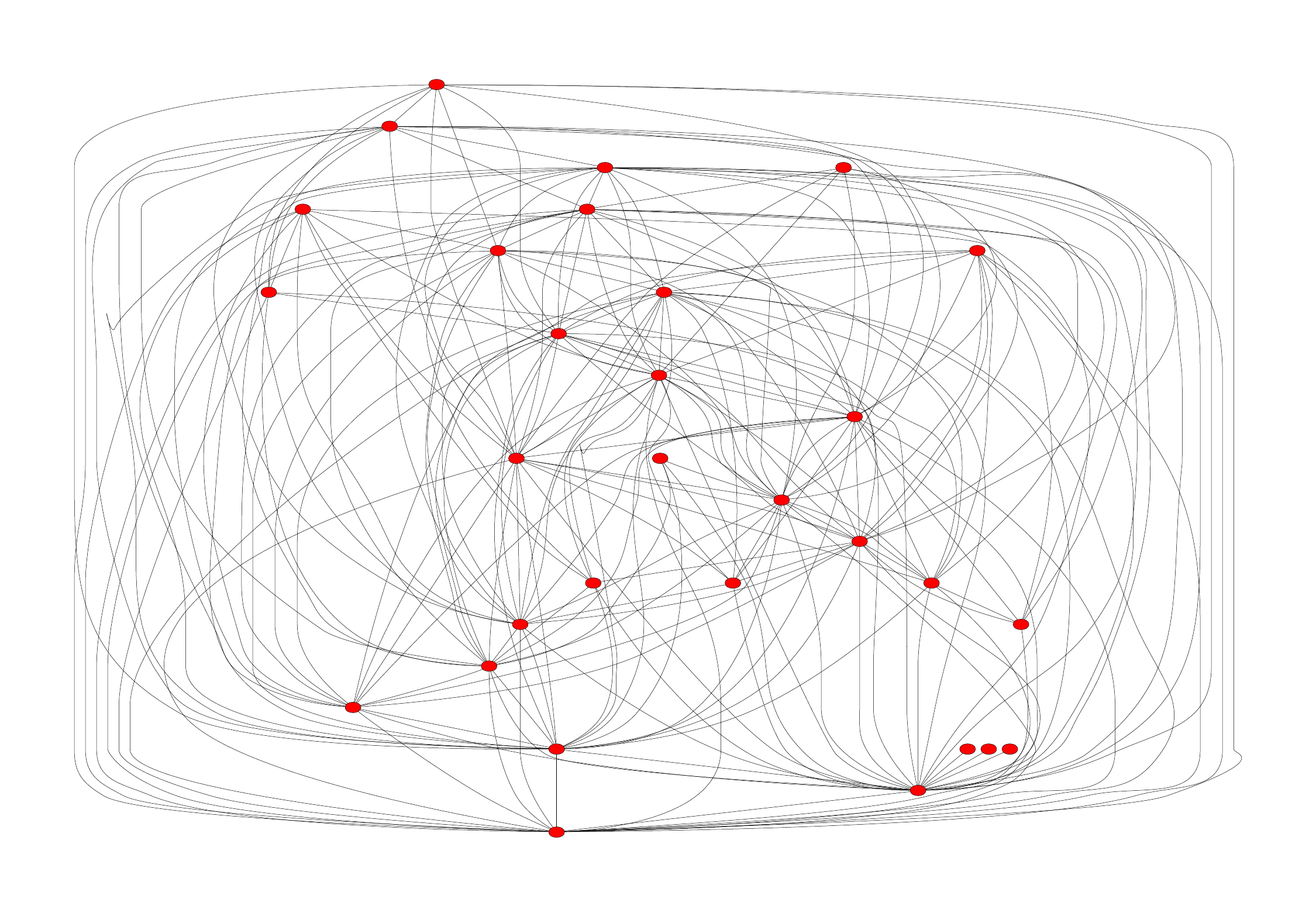}
    \caption{A sampled social graph from a real-world organization.}
    \label{fig:EnterpriseGraph}
\end{figure}
With preliminaries defined, we model the organizational communication through a graph structure. Figure~\ref{fig:EnterpriseGraph} shows a typical enterprise communication graph. Having the communication modeled, we then describe the confidentiality requirements for the model.
\subsection{Organizational Model}
We model the network of organizational communications by a graph with the following properties: 
\begin{itemize}
    \item The graph displays the communications in the organization.
    \item Nodes are the individuals of the organization.
    \item There is an edge between two nodes if one of the nodes communicated with (e.g.\ sent an email to) the other one; the graph is not directed.
    \item Edges may be labeled with a set of zero or more properties. For example, an edge may be labeled with the token ``acquisition'' if the correspondents discussed the topic ``acquisition''. 
    \item The target application is to perform queries over the graph to measure statistics of edge properties.
\end{itemize}

In principle, ``queries over the graph to measure statistics of edge properties'' may imply training a language model or performing similar natural language tasks over the graph.  For the purposes of our work we focus on a simpler task, which is to safely release a set of edge properties present in the graph. For language tasks, this can be defined as extracting common n-grams from correspondence. Other application scenarios have examined popular item sampling. This problem has been explored in related work like differentially private set union (DPSU) \cite{Gopi2020}, and top-$k$ item selection \cite{durfee-top-k}.

\subsection{Confidentiality Requirements}
In order to achieve our goal of organizational confidentiality, we impose the following additional requirements:
\begin{itemize}
    \item The privacy mechanism should protect edges (provide edge-level privacy),
    \item the mechanism should account for correlation between neighboring edges,
    \item the mechanism should protect against changes to edge properties, but not to changes in graph structure (the graph edges are considered invariant and public, whereas the properties of the edges are private\footnote{For example, an organization's leadership structure is usually public knowledge, and individuals with internal access can usually determine who reports to who and infer common lines of communication.}),
    \item the graph structure is known to attackers, and
    \item attackers may have access to a data generation model $\theta$ that can predict an edge's properties, given its neighbors'. 
\end{itemize}
Note that the data generation model $\theta$  assumes that neighboring edges influence one-another but non-neighboring edges don't.   That is, an edge's properties are conditionally independent of the rest of the graph, given its neighbors' properties. This is equivalent to expressing the graph as a Markov random field. In practical scenarios, edge correlation may be effective beyond the neighborhood of an edge. However, we believe our conditional independence assumption is a reasonable approximation.

The reader will note that this problem framing lends itself well to Pufferfish privacy: the presence or absence of a property on an edge can be framed as a pair of complementary secrets $s^0_i$ and $s^1_i$ respectively for all edges $X_i$ in the graph. In the Pufferfish instantiation this leads to having the set of secrets
$\mathcal{S} = \{s^0_i, s^1_i : \textnormal{for all edges $i$ in the graph}\}$, so the status of the corresponding property of each edge is a secret. The set of secret pairs becomes $\mathcal{S}_{\textnormal{pairs}} = \{(s^0_i, s^1_i) : \textnormal{for all edges $i$ in the graph}\}$ as we desire that the adversary cannot tell if each edge has the property of interest or not. Finally, attackers have access to generating distributions in $\Theta$ describing how properties may be defined on the graph, and in particular how they may be correlated.

In the following section we describe how we can leverage Pufferfish privacy to protect edge properties, while accounting for neighborhood correlation, and subsequently we present empirical results on a language task applied to a real-world organizational graph.

\section{Mechanism Design} \label{PrivMech}
We consider the graph representation of the organizational communications, consisting of nodes for individuals and edges for the correspondences among them. 
\subsection{Neighborhood model for correlation} \label{Sec: NeighborCorr}

\begin{figure}[ht]
    \centering
    \includegraphics[scale=0.35]{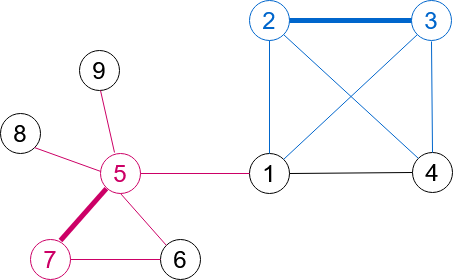}
    \caption{Neighborhood correlation, each edge is correlated with its adjacent edges. The 23 edge is adjacent to edges 21, 24, 31 and 34. Similarly, edge 57 is adjacent to edges 51, 56, 58, 59, and 76.}
    \label{fig:NeighborCorr}
\end{figure}

In our neighborhood model, we capture the correlation among the \emph{adjacent edges} as shown in Figure \ref{fig:NeighborCorr}. This is one choice of modeling correlation for the communication between nodes and one can think of other models such as the clique model in the flu status over social network example of \cite{song2017} where the network is a union of cliques and each clique has a dependency among its nodes. 
This may be appealing for the organization communications considering each group in the organization as a clique. However, in practice it is seldom the case that every pair of individuals in a particular group has a communication link between them, and therefore the actual cliques in the communication graph correspond to only subsets of the groups in the organization, not capturing the correlation effectively. Furthermore, the set of maximal cliques in a large graph may be prohibitively expensive to enumerate and estimate $W_\infty$ for each. In our model we instead define the graph as a union of neighborhoods, where each neighborhood is defined as a central edge and its adjacent neighbors.  Note that the conditional independence assumption implies that knowledge of the edge properties in the neighborhood is sufficient to determine the properties of the central edge.

A change in an edge's properties will influence its neighbors. For example, an edge labeled with the property ``acquisition'' may imply that neighbors are much more likely to share this property.  If we can model the effect of this change probabilistically, we can compute the Wasserstein distance between query distributions, providing a sensitivity measure that accounts for an edge's correlation with its neighbors. In the next section we outline our privacy definition, followed by three proposed models for capturing neighborhood correlation.


\subsection{Privacy Definition}
We use  Pufferfish privacy to design a private mechanism that takes correlation into account. We start with designing a private mechanism for counting one property in the graph, and then extend it to all property counts: 
\begin{enumerate}
    \item The database is a set of records: $X=\{X_1, \cdots , X_N\}$; $X_i= 0$  or $X_i= 1$ corresponding to the events the edge \emph{i} has the property or not\footnote{Or: whether the property frequency is above a certain threshold or not.}, indicating complementary secrets $s_i^0$ or $s_i^1$. 
    \item The Pufferfish parameters: $(\mathcal{S}, \mathcal{S}_{\textnormal{pairs}}, \Theta)$: the set of secrets $\mathcal{S} = \{s_i^0, s_i^1; i= 1, \ldots, n\}$, the secret pairs to be indistinguishable $\mathcal{S}_{\textnormal{pairs}} = \{(s_i^0, s_i^1), i=1, \ldots, n\}$, and $\Theta$, the set of models describing the correlation. The Pufferfish privacy guarantee is shown in Equation \eqref{eq: Pufferfish}. 
    \begin{equation}
    \label{eq: Pufferfish}
    e^{-\epsilon} \leq \frac{\Pr_{M,\theta}(M(X)=w|s_i^0,\theta)}{\Pr_{M,\theta}(M(X)=w|s_i^1,\theta)} \leq e^{\epsilon}
    \end{equation}
    \begin{enumerate}
        \item In the organizational communication $\mathcal{S}$ consists of the binary values of each $X_i$, $i = 1, \ldots, n$.
        \item $\mathcal{S}_{\textnormal{pairs}}$ is what we want the property label to be indistinguishable from. Since the label is binary, $\mathcal{S}_{\textnormal{pairs}}$ is $(s_i^0, s_i^1)$, which indicate the existence or non-existence of an property in an edge. 
        \item $\Theta$: Instead of considering a set of correlations, we focus on the neighborhood correlation for $\theta \in \Theta$. We use the Markov Quilt mechanism from~\cite{song2017} for this model. However, unlike the Markov Quilt mechanism, we rely on empirical data and measure the exact correlation inside the quilt. 
        \item Query $f$: Maps the dataset $X$ into a scalar $f(X) = |i \in \{1, \ldots, n\} : X_i = 1|$, counting the number of edges having the property of interest.
    \end{enumerate}
    \item Markov Quilt. 
    We adapt the Markov Quilt Mechanism to protect the edges in our neighborhood model. We assume that each edge $X_i$ is correlated to its adjacent edges $X_N$ and has no correlation with the rest of the graph (neither to $X_R$, nor to $X_Q$), i.e. $\delta =0$. The $\textnormal{card}(X_N)$ 
    translates to the maximum number of adjacent edges to an edge, i.e. $2 \times D_{max} - 1 $, where $D_{max}$ is the maximum degree of a node in the graph, and we subtract 1 so as not to double-count the central edge itself. We note that
    the application of $2 \times D_{max}$ group differential privacy (Corollary \ref{group_DP}) would be a baseline for our case.
    
\end{enumerate}


\subsection{Correlation Models} \label{correlation-models}
In order to assess the Wasserstein distance between neighboring secret pairs (changing a single property from true to false or vice-versa), we require a model that can estimate by how much a neighborhood's labels might change due to a change in the central edge's label. 

Specifically, we seek to estimate $\Pr(f(X) = w | s_i, \theta)$.  The attacker's priors, encoded by $\theta$ indicate to what accuracy the attacker may be able to estimate the change in $f(X)$ if edge $X_i$ changes its label, replacing $s_i^0$ with $s_i^1$ or vice versa. 
By applying the Markov assumption, we need only measure the impact of a label change on an edge's immediate neighborhood (i.e.\ its impact on the Markov quilt).  Thus, for each of the correlation models below, $w$ is measured for the local neighborhood and the rest of the graph is assumed to be constant.

\subsubsection{Conditional Model}
Our first model, namely the \textbf{Conditional} model  estimates the probability $\Pr(f({X})=w | s_i, \textnormal{deg}(X_i), \textnormal{freq}(a))$, where $\textnormal{deg}(X_i)$ is the number of edges adjacent to edge $X_i$, $\textnormal{freq}(a)$ is the attacker's prior on the frequency of property $a$ (ie their prior estimate of how many edges are labeled with $a$, perhaps sampled from a public corpus).  We construct this model empirically by bucketizing $\textnormal{deg}(X_i)$ and $\textnormal{freq}(a)$ on a logarithmic scale, sampling up to 100 edges per bucket, and building a histogram describing $\Pr(f({X})=w)$, for discrete intervals of $w$. Note that since the rest of the graph is invariant to changes in $s_i$, we need only measure this distribution over values of $w$ specific to the edge's local neighborhood. This is our highest-fidelity model and represents the most knowledge we assume an attacker may possess about the graph.

\subsubsection{Global Model}
Our second model, called the \textbf{Global} model ignores $\textnormal{deg}(X_i)$ and $\textnormal{freq}(a)$ and empirically measures $\Pr(f({X})=w | s_i^j)$ for secrets $s_i^0$ and $s_i^1$. The resulting model is a normalized frequency histogram of how often $f({X})=w$ when the central edge's property is set, $s_i^1$ , and a separate histogram for when it is not set $s_i^0$.  As in the conditional model, we measure the distribution over the range of values that $w$ may take over a local neighborhood, assuming the rest of the graph to be constant.

\subsubsection{Binomial Model}
Finally our third model, which we call the \textbf{Binomial} model, representing the least amount of attacker knowledge, empirically estimates $p_i=\Pr(s_j|s_i)$, the probability distribution over a randomly selected adjacent edge's secrets, given the label of the central edge, and then estimates $\Pr(f({X})=w|s_i)$ as a Binomial distribution parameterized by $p_i$ and $\textnormal{deg}(X_i)$:
\[
P(f({X})=w|s_i) = \textnormal{Binomial}(\textnormal{deg}(X_i), p_i)
\]

That is, for a given edge $X_i$ with $\textnormal{deg}(X_i)$ neighboring edges, if $X_i$'s label is $s_i$, then the probability that it has $w$ neighboring edges with a true label is represented by the distribution of successes over $\textnormal{deg}(X_i)$ Bernoulli trials with success probability $p_i$.  Note that $p_i$ is considered a constant over the whole graph, independent of $\textnormal{deg}(X_i)$, and is parameterized by the central edge's label $s_i$.

\subsection{Measuring $W_{\infty}$}

\begin{figure}
    \centering
    \includegraphics[scale=0.47]{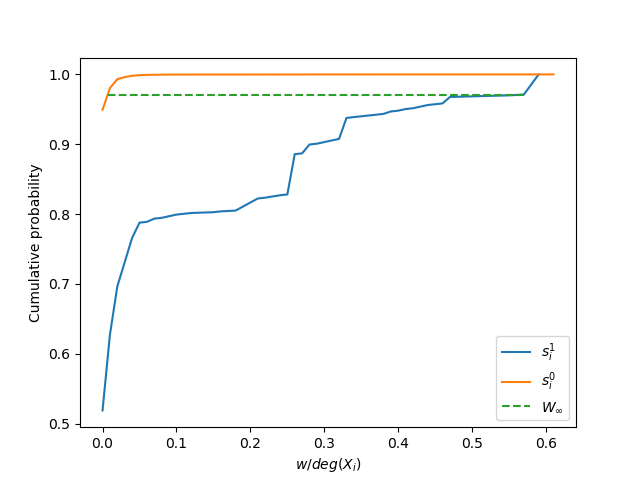}
    \caption{Example $W_\infty$ determined by differencing cumulative distributions for the \textbf{Conditional} model. } 
    \label{fig:wass}
\end{figure}

Our correlation models afford a straightforward estimation of the maximal Wasserstein distance $W= max_{X_i\in X} W_\infty(X_i)$.  For each neighborhood in the graph, instantiate distributions $\Pr(f({X})|s_i^0)$ and $\Pr(f({X})|s_i^1)$, and measure  $W_\infty$.  Computationally, this is equivalent to determining the largest horizontal distance between the  two distributions when they are expressed as cumulative functions, as illustrated in Figure~\ref{fig:wass}.  $W$ is then the maximal $W_\infty$ over all edges. 

Note that $W$ is bounded above by the largest neighborhood size: flipping a single edge property may trigger a flip in at most $\textnormal{deg}(X_i)$ adjacent edges.

\section{Experiments} \label{experiments}


We run our experiments on the Avocado corpus\cite{avocado2015}.  In order to accurately recognize messages sent to multiple recipients, we apply a simple inference heuristic for identifying organizational mailing lists and their memberships (enabling, for example, the expansion of ``All Employees'', instantiating an edge from sender to each member of this list. 
The complete graph contains 393 nodes (individuals) and 21312 edges representing correspondence between users.  The largest neighborhood in the graph consists of 1883 edges.

Edges are subsequently labeled with properties.  We extract unigrams and bi-grams from messages passed between edges and set the edge property $X^a_i$ to ``true'' for each ngram $a$.  Thus an edge with the property ``acquisition'' set to true indicates that at least one message passed between the connected nodes containing the word ``acquisition''.  Edges with no such property are implicitly ``false'' for that property.

\def\boxit#1{%
  \smash{\fboxsep=0pt\llap{\rlap{\fbox{\strut\makebox[#1]{}}}~}}\ignorespaces
}
\definecolor{Gray}{gray}{0.91}
\begin{table}[t]
    \centering
    \begin{tabular}{|c|c|c|c|}
        \toprule
        \rowcolor{Gray}
         $\log(\textit{freq})$	&$\log(\textit{deg})$	&$W_{\infty}$&	$W$ \\ \midrule
0&	0&	1.0	& 10.0\\
\rowcolor{Gray}
0&	1&	0.08&	8.0\\
0&	2&	0.02 &	20.0\\
\rowcolor{Gray}
0&	3&	0.01 & 18.83\\
1&	0&	1.0&	10.0\\
\rowcolor{Gray}
1&	1&	0.58&	57.0 \\
1&	2&	0.5&	500.0\\
\rowcolor{Gray}
1&	3&	0.09&	169.47\\
2&	0&	0.71&	7.1\\
\rowcolor{Gray}
2&	1&	0.66 &	66.0 \\
2&	2&	0.74&	740.0\\
\rowcolor{Gray}
 \hspace{-0.3in}\boxit{2.1in}\hspace{0.3in}  2 &	 3&	 0.51 &	 960.33  \\
3&	0&	0.5 &	5.0 \\
\rowcolor{Gray}
3&	1&	0.31 &	31 \\
3&	2&	0.37&	370.0\\
\rowcolor{Gray}
3&	3&	0.29 &	546.07 \\
4&	0&	0.36 &	3.6 \\
\rowcolor{Gray}
4&	1&	0.16&	16.0\\
4&	2&	0.21&	210.0 \\
\rowcolor{Gray}
4&	3&	0.13 &	244.79\\
\bottomrule
    \end{tabular}
    \caption{Wasserstein metrics for edges with neighborhoods of size $\textit{deg}$ and properties with global frequency $\textit{freq}$. The boxed row represents the highest sensitivity.}
    \label{tab:model1}
\end{table}

With edge properties set, we construct the three correlation models, as described in section~\ref{correlation-models}.  Table~\ref{tab:model1} shows the estimated Wasserstein measures for the various property frequencies and neighborhood size under the \textbf{Conditional} correlation model.  The maximum influence $W_\infty$ 
of a bucket is scaled by the maximum neighborhood size for the bucket, up to the largest possible neighborhood in the graph $N_{max}=1883$.  That is:
\[
W=W_\infty*\min(N_{\textit{max}},10^{\log(\textit{deg})+1})
\]



We note several points of interest in Table~\ref{tab:model1}.  First, that while the largest $W$ corresponds to large neighborhoods ($\log(\textit{deg})=3$), it doesn't necessarily correspond with high-frequency or low-frequency properties. Second, we note that the maximal $W$ is roughly equivalent to half the largest neighborhood, achieving some improved utility over group privacy.

\begin{figure}[ht]
    \centering
    \begin{subfigure}
    \centering
    \includegraphics[scale=0.47]{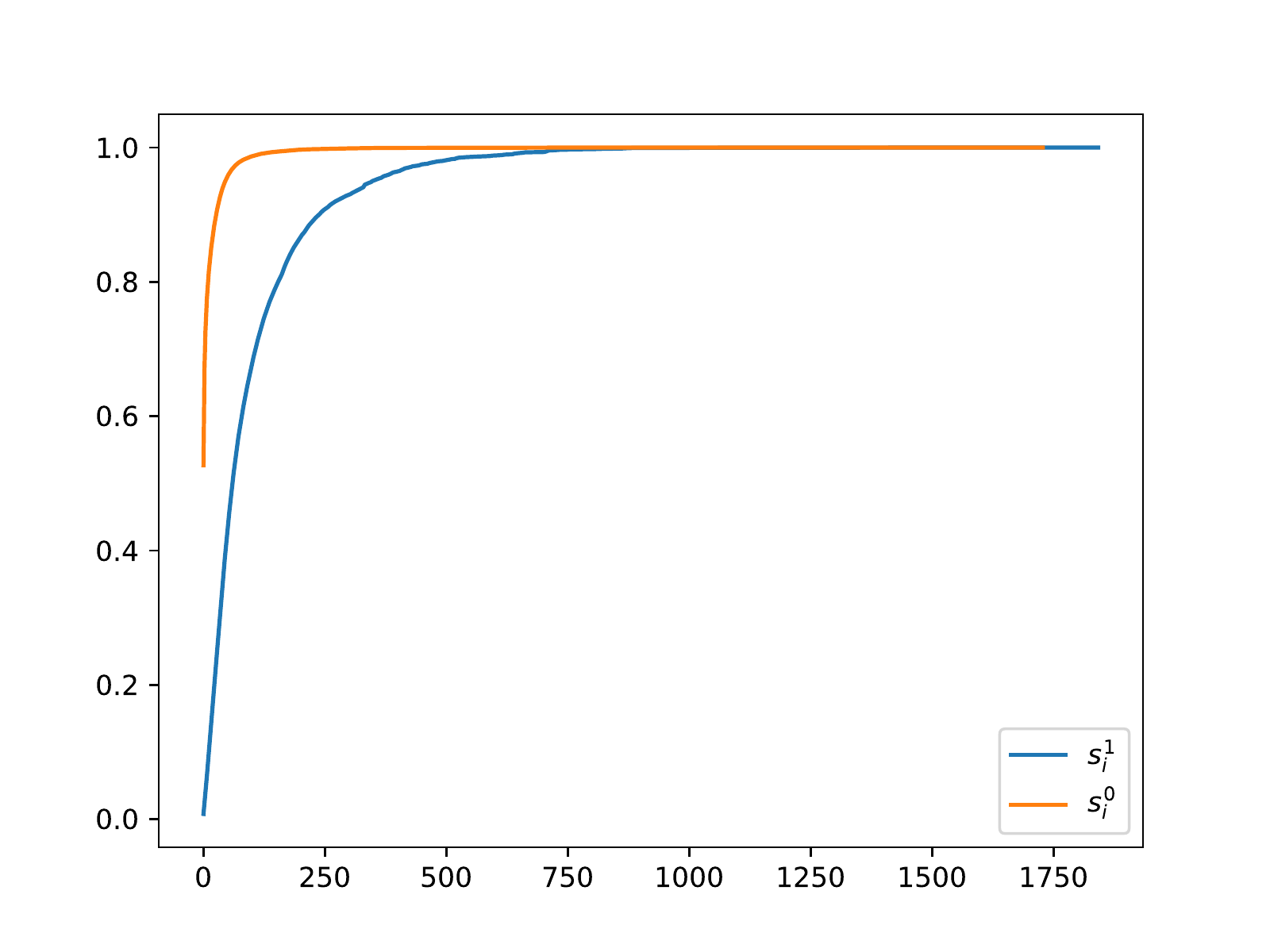}
    \caption{Cumulative distributions of $\Pr(f(X)=w|s_i^j)$ for the \textbf{Global} model, conditioned on secret $s_i^j$. }
    \label{fig:cdf_bernoulli}
    \end{subfigure} %
    \begin{subfigure}
    \centering
    \includegraphics[scale=0.47]{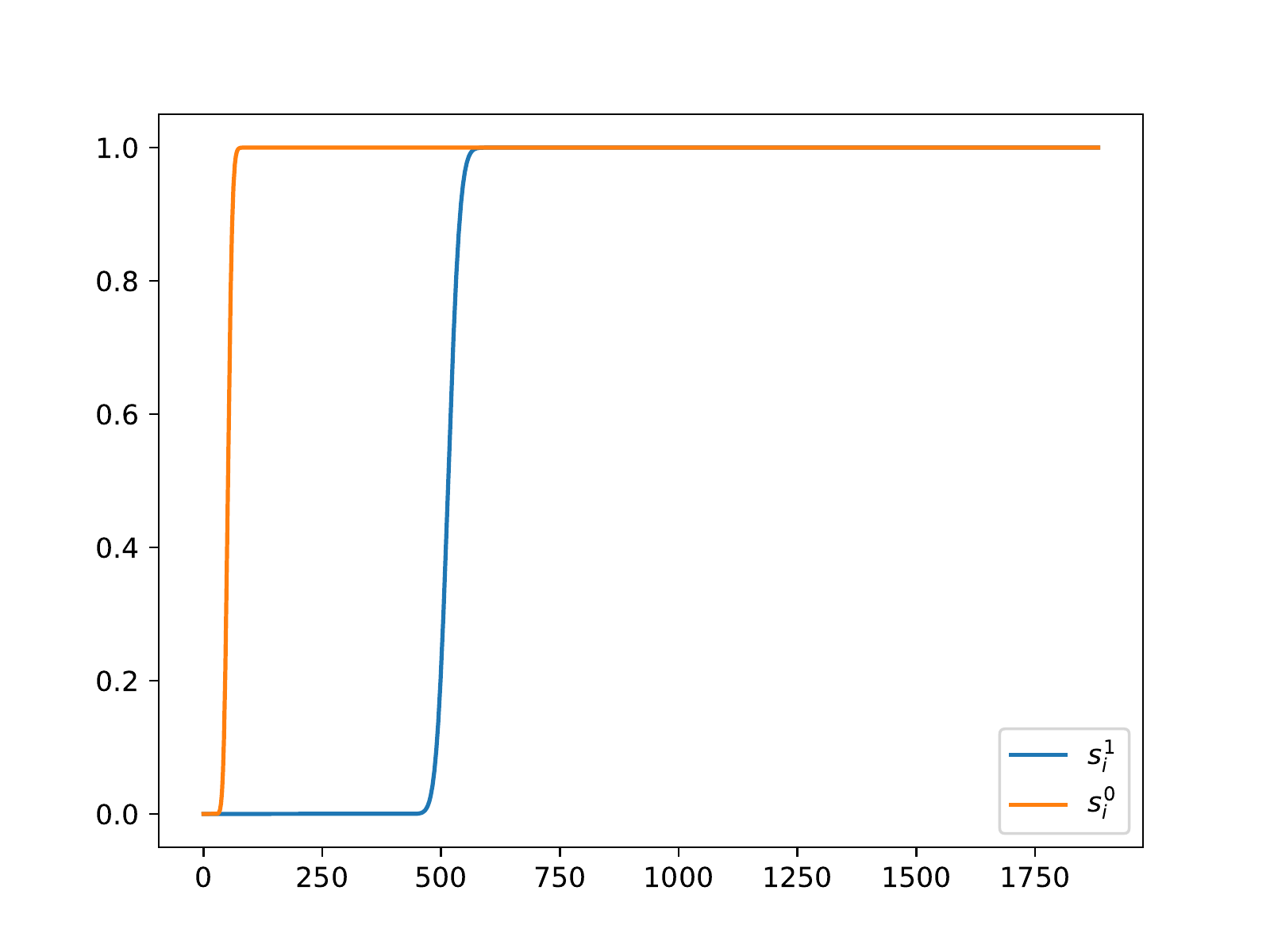}
    \caption{Cumulative distributions of $\Pr(f(X)=w)$ assuming binomially distributed  counts for secrets $s_k^j$ on edges $k$ adjacent to $X_i$.}
    \label{fig:binom}
    \end{subfigure}

\end{figure}

Our second model, the \textbf{Global} model measures $\Pr(f(X)=w | s_i)$ directly for secrets $s_i^0$ and $s_i^1$. The cumulative distribution functions of these measures are shown in Figure~\ref{fig:cdf_bernoulli}.  The maximal Wasserstein measure is the maximum horizontal distance between these two distributions, or 866.  

Finally, the \textbf{Binomial} model represents the two label distributions by estimating Bernoulli parameters $p_0$ and $p_1$ for each label respectively, and measuring the maximal Wasserstein distance between these distributions $\Pr(f(X)=w | s_i^j)=\textnormal{Binomial}(\textnormal{deg}(X_i), p_j)$. Using this approach we emprically measure $p_0$ to be  0.0277 and $p_1$ to be 0.2739.  These parameters indicate that an adjacent edge is about ten times more likely to have property $a$ if the central edge has property $a$.

Figure~\ref{fig:binom} shows the binomial distributions for these Bernoulli parameters, and again the maximal Wasserstein measure is the maximum horizontal distance between the curves, or 558.  Compared with the \textbf{Global} model, we observe that choosing a binomial distribution is a relatively poor approximation of the empirical behavior, but may represent an attacker's best guess as to how neighborhoods vary when edge properties change.


\subsection{Language Tasks}
\begin{table*}[t]
    \centering
    \begin{tabular}{|c|c|c|c|c|}
\toprule 
\rowcolor{Gray}
Description &$W$ &  $\lambda$ &$\texttt{yield (\%)}$ & RMSE   \\
\midrule
Edge-level privacy & 1 & 10.0 & 809200.5 $\pm$ 529.6 (58.2\%) & 12.0 $\pm$ 0.02  \\
\rowcolor{Gray}
Node privacy   & 1 & 10.0 &  77679.0 $\pm$ 159.4 (5.6\%) & 6.4 $\pm$ 0.06 \\
Binomial Model  & 558 & 5580.0 &  695823.0 $\pm$ 521.3 (50.1\%) & 5577.7 $\pm$ 8.14 \\
\rowcolor{Gray}
Global Model  & 866 & 8660.0  &  695316.8 $\pm$ 647.3 (50.0\%) & 8665.8 $\pm$ 12.15  \\
Conditional model  & 960 & 9600.0 & 695366.3 $\pm$ 561.83  (50.0\%) & 9594.2 $\pm$ 12.24  \\
\rowcolor{Gray}
Group privacy  & 1883 & 18830.0 &   695195.5 $\pm$ 428.6 (50.0\%) &  18833.3 $\pm$ 26.73 \\
\bottomrule

    \end{tabular}
    \caption{Experimental results for histogram publication, $\epsilon=100$. We measure yield- the number of positive ngram counts for each of the correlation models, as well as for node-level, edge-level, and group privacy.}
    \label{tab:hist}
\end{table*}

\begin{table}[ht]
    \centering
    \begin{tabular}{|c|c|c|c|}
\toprule
\rowcolor{Gray}
Description& $W$ & $E[\texttt{yield}]$ & $\sigma$  \\
\midrule
Edge-level privacy &  1 & 24814.9 & 46.3 \\
\rowcolor{Gray}
Node privacy  &  1   & 91 & 3.35\\
Binomial Model & 558  & 228.7 & 5.71  \\
\rowcolor{Gray}
Global Model & 866  & 135 & 6.51  \\
Conditional model  & 960  & 116.7 & 7.29 \\
\rowcolor{Gray}
Group privacy  & 1883 & 41.9 &  3.11  \\
\bottomrule
    \end{tabular}
    \caption{Experimental results for DPSU, $\epsilon=100$. We measure yield- the number of extracted ngrams for each of the correlation models, as well as for node-level, edge-level, and group privacy.}
    \label{tab:dpsu}
\end{table}

We apply the privacy mechanism to two query tasks, histogram release, and differentially private set union (DPSU)~\cite{Gopi2020}.

\subsubsection{Histogram release}
Our first task involves generating a histogram over edge properties. When edge properties are ngrams, the task is equivalent to computing the frequencies of ngrams in the corpus.  To limit the sensitivity of the histogram to changes in a single edge, we limit the contribution of each edge to $c=1000$ distinct ngrams. The value of $c$ determines the maximum number of ngrams each edge contributes and we choose the $c$ most common on each edge. Note that for this task it is assumed that the domain of ngrams is known \emph{a priori} from a public source.  In practice it is usually necessary to identify these using the private corpus as well, which we address in the second experiment.   

Histogram publication 
can be accomplished using the Laplace mechanism, adding Laplace noise with scale parameter $\lambda=cW/\epsilon$, where $c$ is the per-edge contribution limit, $W$ is the maximal Wasserstein distance accounting for edge-neighborhood correlation, and $\epsilon$ is the privacy budget. 

For the purposes of this experiment we choose a relatively large $\epsilon=100$, as the corpus is comparatively small for running effective privacy mechanisms. It has been noted in other work that privacy mechanisms on graphs require large values of $\epsilon$, e.g.~\cite{Chen2014}.  To measure the utility of the result, we assess root mean squared error (RMSE) between the noisy and true histogram, and also indicate the ``yield'' of the histogram, measured as the number of ngrams with positive counts (note that some noisy counts may be negative).  Positive counts are necessary for language modeling tasks such as computing inverse-document-frequency~\cite{jones1972statistical}, and a large number of negative counts indicates lower utility of the resulting histogram. Results are reported over ten trials.

Table~\ref{tab:hist} contains the results for the histogram task.  Note that even with a large epsilon, and a moderate amount of noise, in the best case (edge-level privacy) only 58\% of ngrams have useful counts. However, unlike the Conditional, Global, or Binomial models this result doesn't account for correlation. Note that in the case of group privacy the added noise is comparable to the total number of edges in the graph. 

\subsubsection{DP Set Union Application}
In the histogram publication experiment we noted that it is assumed that the domain of ngrams is known \emph{a priori}, an assumption which may not be true. For instance, real-world language modeling applications may call for differentially private vocabulary selection. Differentially private set union (DPSU) aims to identify the union of elements in $k$ input sets (in our setting, sets of edge properties on $k$ edges).  This problem was addressed in~\cite{Gopi2020}. To account for edge correlation, as in the case of histogram publication, it is necessary to scale the sensitivity of the property counts by $cW$, as changing any edge can change as many as $c$ properties and may influence its neighborhood by a factor as large as $W$. For this experiment we compare the yield (the number of published n-grams) of the privacy mechanism over ten independent applications of the mechanism, for each of the three correlation models.  We also provide baseline yield for node-level, edge-level, and group privacy.  We choose $\epsilon=100$ and $c=1000$ as in the previous experiment.   

The results of this experiment are shown in Table~\ref{tab:dpsu}.  As in the previous experiment, the best result corresponds to edge-level privacy, which neglects to account for edge-neighborhood correlation.  Of the approaches that address correlation, the binomial model yields the largest set of ngrams. These experimental results illustrate the challenges associated with differentially private language modeling tasks.  If $\epsilon$ is large, and the number of private entities numbers in the tens of thousands, the yield of a DP mechanism can still be very limited. Despite these observations, our results illustrate how a privacy mechanism can be appropriately modified to account for neighborhood correlations in the graph, and we believe algorithmic yields can only improve with larger graphs.


\section{Conclusion}

Natural language tasks are an area of increasing relevance with recent advances in language modeling capabilities, and the availability of large language corpora. In order to train language models targeting organizations, it is necessary to address latent confidentiality concerns in the data-- even models that are released for internal use may leak information across groups of users.  In this paper we explored the problem of preserving organizational confidentiality in language tasks, leveraging the Pufferfish privacy framework, while addressing non-IID data among edges in the organization's social network.

We showed how our proposed scheme presents a compromise between two extreme measures of privacy, record-level differential privacy and group privacy. By taking record correlation into account, our scheme provides a more meaningful notion of privacy than record-level differential privacy, while improving the utility compared to group privacy.   

In order to address correlation in the graph, we imposed a Markov assumption enabling us to locally model the sensitivity of graph queries under changes to edge properties, and estimated a maximal Wasserstein metric under varying assumptions about attackers' priors. To demonstrate our approach we applied the mechanism to a real-world organizational dataset and measured performance on a pair of statistical graph query tasks.

Queries over n-grams are only a small part of developing language models from natural language data, and our work leaves many open questions with respect to the development of language models with high utility. For example, can the Wasserstein mechanism be extended to model training tasks, such as the application of stochastic gradient descent~\cite{abadi2016deep}?  

Furthermore, we model correlation up to and including very large neighborhoods of edges, raising the question of how large neighborhoods in a closed graph should be treated-- is there value in preserving confidentiality within the organization once neighborhoods reach some critical size?  Clearly, we can derive improved utility if we are not required to protect these groups.  Our future work aims to address these and related questions.























\bibliographystyle{plain}
\bibliography{PrivacyInEnterprise}
\end{document}